\documentstyle[12pt]{article}
\begin{document}

\author{A. de Souza Dutra(1)\thanks{%
dutra@feg.unesp.br} and C. P. Natividade (1,2) \\
(1) UNESP/Campus de Guaratinguet\'a-DFQ\\
Av. Dr. Ariberto Pereira da Cunha, 333\\
Guaratinguet\'a - SP - Brasil\\
CEP 12500-000\\
(2) Instituto de F\'{\i}sica - Universidade Federal Fluminense\\
Av. Litor\^{a}nea S/N, Boa Viagem, Niter\'{o}i\\
Rio de Janeiro - RJ - Brasil\\
CEP 29210-340}
\title{\bf Class of self-dual models in three dimensions.}
\maketitle

\begin{abstract}
In the present paper we introduce a hierarquical class of self-dual models
in three dimensions, inspired in the original self-dual theory of
Towsend-Pilch-Nieuwenhuizen. The basic strategy is to explore the powerful
property of the duality transformations in order to generate a new field.
The generalized propagator can be written in terms of the primitive one
(first order), and also the respective order and disorder correlation
functions. Some conclusions about the ``charge screening'' and magnetic flux
were established.
\end{abstract}

From the mathematical point of view, topological theories in three
dimensions contains a rich variety of models which been received much
attention in the last years. One of them is the self-dual model \cite{sd},
which presents a close connection with the well established Chern-Simons
theory \cite{chern-simons}. This fact could be confirmed by different ways,
for instance, by comparing the Green functions of the Maxwell Chern Simons
(MCS) theory and Self-dual (SD) model \cite{chern-simons}\cite{banerjee}, by
inspecting the constraint structure of each model \cite{banerjee} or through
the bosonization of the massive Thirring model in three dimensions, which is
related to the MCS theory in the large mass limit \cite{fradkin}. In this
last case, the equivalence between both models has been obtained starting
from a careful analysis of the partition function and was improved later,
through the calculation of higher order derivative terms \cite{fermion-boson}%
.

In the present work, we shall introduce a hierarchcal family of dual models
in three dimensions, related to the original SD model. The mathematical
structure of the SD theory offers an alternative way of building up $N$
families of dual models. At the final step, it is generated a master
Lagrangian density corresponding to a higher order derivative model. A very
interesting aspect of this model is the existence of an isomorphism between
its observables and those obtained in its first order form. This fact can be
proved through different procedures. Firstly, in the canonical analysis of
the fields and their momenta, by using the treatment of order reduction \cite
{rubens}. In what follows, we will use a method developed in a series of
papers \cite{marino} (see also \cite{others} for related works), in order to
describe the magnetic flux and charge on the plane ($x^{1},\,x^{2}$) through
two dual operators ($\mu ,\,\sigma $), called disorder and order operators
respectively.

In order to implement our alternative model, let us begin exploring the
mathematical structure of the self-dual fields. In this sense, let us
consider the duality transformation of the primary field $A_{\mu }^{\left(
N\right) }$, 
\begin{equation}
A_{\mu }^{\left( N\right) }\,=\,\epsilon _{\mu \alpha \beta }\partial
^{\alpha }A_{\beta }^{\left( N+1\right) },  \label{eq1}
\end{equation}

\noindent where the index $N$ is an integer which identifies the family of
the respective self-dual field. The relation (1) gives rise to the
possibility of generating a class of Lagrangian densities indexed by $N$.

Let us start our study by considering the following Lagrangian density, 
\begin{equation}
L_{1}\,=\,-\frac{a}{4}\left( F_{\mu \nu }\right) ^{2}+\,b\,\partial _{\mu
}F^{\mu \lambda }\,\partial ^{\nu }F_{\nu \lambda }\,+\,\theta \,\epsilon
^{\mu \nu \rho }\,\partial _{\sigma }\,A_{\mu }\,\partial _{\nu }\partial
^{\sigma }\,A_{\rho },  \label{eq2}
\end{equation}

\noindent which has been examined recently \cite{csco}, with $a$, $b$ and $%
\theta $ defined in it. Now, we are going to show that the Lagrangian
density appearing in (2) is a higher order extension from the
Proca-Chern-Simons one: 
\begin{equation}
L_{1}^{\left( 0\right) }\,=\,-\frac{a}{2}A_{\mu }^{\left( 0\right) }\,A^{\mu
\left( 0\right) }\,+\,\frac{b}{2}\left( F_{\mu \nu }^{\left( 0\right)
}\right) ^{2}\,+\,\theta \,\epsilon ^{\mu \nu \rho }\,\,A_{\mu }^{\left(
0\right) }\,\partial _{\nu }\,A_{\rho }^{\left( 0\right) }.  \label{eq3}
\end{equation}

\noindent By using the transformation (1), with $N=0$, it is lengthy but
straightforward to show that we arrive at the Lagrangian density (2). The
propagators can be related among them, since $<A_{\mu }^{\left( 0\right)
}A_{\nu }^{\left( 0\right) }>\,=\,-\,\epsilon _{\mu \alpha \beta }\,\epsilon
_{\nu \grave{\alpha}\grave{\beta}}\,<\partial ^{\alpha }A_{\beta }^{\left(
1\right) }\partial ^{\grave{\alpha}}A_{\grave{\beta}}^{\left( 1\right) }>$.

From the above considerations becomes clear that the results obtained here
can be generalized from the $N$-order to the ($N+1$) one. Therefore, from
the basic Lagrangian density given by Eq.(\ref{eq3}), we can build up the
following generic higher-order Lagrangian density: 
\begin{equation}
L^{\left( N\right) }\,=\,\frac{\left( -1\right) ^{\left( N-1\right) }}{4}%
F_{\mu \nu }^{\left( N\right) }\Box ^{\left( N-1\right) }\left( a\Box
+b\right) F^{\mu \nu \left( N\right) }\,-\,\left( -1\right) ^{\left(
N-1\right) }\theta \,\,\epsilon ^{\mu \nu \rho }\,\,A_{\mu }^{\left(
N\right) }\,\partial _{\nu }\,\Box ^{\left( N\right) }\,A_{\rho }^{\left(
N\right) }.  \label{eq5}
\end{equation}

The above Lagrangian density belongs to a class such that the first one is
related to the bosonization of the massive Thirring model \cite
{fermion-boson}. In order to simplify the calculation of the canonical
momenta, we are going to define the quantities 
\begin{equation}
f_{\mu }\,\equiv \,\sqrt{\Box ^{N-1}}\,A_{\mu }\,,\,\,\dot{f}\,_{\mu
}\,\equiv \,\sqrt{\Box ^{N-1}}\,\dot{A}_{\mu }\,,  \label{eq6}
\end{equation}

\noindent where $\Box ^{-n}=\int \frac{d^{3}k}{\left( 2\pi \right) ^{3}}%
\left( k^{2}\right) ^{-n}e^{ikx}$. Now, if we take the rescaling $\Box
^{-n}\rightarrow \left( \Box -\Omega \right) ^{-n}$, and take $\Omega
\rightarrow 0$ at the end of the calculations, the expansion in powers of
the d 'Alambertian can be employed by acting on the fields. Consequently, we
can derive the canonical momenta associated to independent variables $\left(
f_{\mu },\,\dot{f}_{\mu }\right) $ in a natural way\thinspace $\delta
S^{\left( n\right) }=\int d^{3}x\,\sum_{n=0}^{\infty }\,\frac{d}{dt}\left(
\pi _{\nu }^{\left( n\right) }\delta f^{\nu }+S_{\nu }^{\left( n\right)
}\delta \dot{f}^{\nu }\right) $, where now the action $S^{\left( n\right)
}=\int dt\,L^{\left( n\right) }$ is the reduced form from those in equation (%
\ref{eq5}). Therefore, the momenta become 
\[
\pi ^{\nu \left( N\right) }\,=\,\left( -1\right) ^{N-1}\left\{ b\,f^{0\nu
\left( N\right) }\,+\,a\left( \partial ^{k}\partial ^{\lambda }f_{\lambda
}^{0\left( N\right) }\delta _{k}^{\nu }\,-\,\partial ^{0}\partial ^{\lambda
}f_{\lambda }^{\nu \left( N\right) }\right) \,-\,2\theta \epsilon ^{\mu
\lambda \nu }\partial _{\lambda }\partial ^{\nu }f_{\mu }\right\} 
\]
\begin{equation}
s^{\nu \left( N\right) }\,=\,\left( -1\right) ^{N-1}2\,a\,\left( \partial
_{\mu }\,f^{\mu \nu \left( N\right) }\,-\,\delta _{0}^{\nu }\partial _{\rho
}f^{0\rho \left( N\right) }\right) \,-\,\theta \,\epsilon ^{0\lambda \nu
}\partial _{0}\,f_{\lambda }^{\left( N\right) },  \label{eq7}
\end{equation}

\noindent which relates physical quantities from $N$-theory with first-order
one. From the above equations we conclude that the basic commutators of the
present theory in the Coulomb gauge are 
\[
\left[ f_{l}^{\left( N\right) }\left( \bar{x}\right) ,\pi _{k}^{\left(
N\right) }(\bar{y})\right] \,_{x^{0}=y^{0}}\,=\,\sqrt{\nabla ^{2\left(
N-1\right) }}\,\left[ A_{l}^{\left( N\right) }\left( \bar{x}\right) ,\,\pi
_{k}^{\left( N\right) }(\bar{y})\right] _{x^{0}=y^{0}}\,=
\]
\begin{equation}
=-i\left( -1\right) ^{N-1}\left\{ b\,\delta _{lk}\delta ^{2}\left( \bar{x}-%
\bar{y}\right) \,+\,\left( b+a\nabla ^{2}\right) \partial _{l}\partial
_{k}G\left( \bar{x}-\bar{y}\right) \right\} 
\end{equation}

\begin{equation}
\left[ \dot{f}_{l}\left( \bar{x}\right) ,\,s_{k}^{\left( N\right) }(\bar{y}%
)\right] _{x^{0}=y^{0}}=-i\left( -1\right) ^{N-1}b\,\delta _{lk}\delta
^{2}\left( \bar{x}-\bar{y}\right) =\sqrt{\nabla ^{2\left( N-1\right) }}%
\left[ \dot{A}_{l}^{\left( N\right) }\left( \bar{x}\right) ,\,s_{k}^{\left(
N\right) }(\bar{y})\right] _{x^{0}=y^{0}},  \label{eq9}
\end{equation}

\noindent and also 
\begin{equation}
\left[ \pi _{l}^{\left( N\right) }\left( \bar{x}\right) ,\,\pi _{k}^{\left(
N\right) }(\bar{y})\right] _{x^{0}=y^{0}}\,=\,-i\,\theta \,\left( -1\right)
^{N-1}\left( b+a\nabla ^{2}\right) \epsilon _{ik}\partial _{l}\partial
^{i}\delta ^{2}\left( \bar{x}-\bar{y}\right) ,  \label{eq10}
\end{equation}

\noindent where $G(\bar{x},\bar{y})$ obeys the equation 
\begin{equation}
\left( b+a\nabla ^{2}\right) \nabla ^{2}G(\bar{x},\bar{y})\,=\,\delta ^{2}(%
\bar{x}-\bar{y}).  \label{eq11}
\end{equation}

\noindent Here we remark that the application of the expansion of $\sqrt{%
\Box ^{N-1}}$ on the above brackets, extract the temporal part of the d
'Alambertian operator.

The Lagrangian density (\ref{eq5}) permit us to infer the corresponding form
of the photon propagator in momentum space 
\begin{equation}
D_{\mu \nu }^{\left( N\right) }(k)\,=\,\frac{-1}{k^{2N}\left(
f^{2}\,-\,4\,\theta ^{2}k^{2}\right) }\left\{ \left( a\,\,+\,\frac{b}{k^{2}}%
\right) P_{\mu \nu }\,+\,2\,i\,\epsilon ^{\mu \lambda \nu }k_{\lambda
}\right\} \,-\,\frac{\xi }{f}\,\frac{k_{\mu }\,k_{\nu }}{k^{2N+2}},
\label{eq12}
\end{equation}

\noindent where the last term corresponds to a gauge fixing. By using
Fourier transform we can obtain the equivalent propagator in the coordinate
space.\ Here, we adopt $P_{\mu \nu }=k^{2}g_{\mu \nu }-k_{\mu }k_{\nu }$ and 
$f\equiv b-a\,k^{2}$. Hence, if we fix some parameters in the original
Lagrangian density given by Eq.(\ref{eq5}) like, $a\equiv 4\,\alpha ^{2}$, $%
b=1$ and $N=1$, we obtain the photon correlation function 
\[
<A_{\mu }(x)\,A_{\nu }(y)>^{dual}\,=\,\left[ \left( 1-4\,\alpha ^{2}\Box
\right) P_{\mu \nu }\,+\,i\,\bar{\theta}\,\epsilon _{\mu \nu \alpha
}\,\partial ^{\alpha }\right] \left( \frac{1}{\left[ \frac{\bar{\theta}}{%
\alpha }\left( 1-\frac{\alpha ^{2}}{\bar{\theta}^{2}}\right) \right] ^{2}}%
\,-\,1\right) \times 
\]
\begin{equation}
\times \,\left( \frac{4\,\alpha ^{2}}{\bar{\theta}^{2}}\right) \,\frac{e^{-%
\frac{\bar{\theta}}{\alpha }\left( 1-\frac{\alpha ^{2}}{\bar{\theta}^{2}}%
\right) R}}{4\,\pi \,R}\,-\,\frac{1}{4\,\pi }\xi \,\partial _{\mu }\partial
_{\nu }\left( 2\,\alpha \,-\,\frac{R}{4}\right)   \label{eq13}
\end{equation}

\noindent with $\bar{\theta}\,\equiv \,i\,\theta $, $4\,\alpha ^{2}\,<\,\bar{%
\theta}^{2}$ and ``dual'' stands for the generalized model defined through
the propagator of the Lagrangian density. The above equation represents the
photon correlation function of the problem mentioned in reference \cite{csco}%
.

At this point, we are able to extract a very interesting and useful result
about the order-disorder correlation functions, starting from equation (\ref
{eq13}). We remember to the reader that the order-disorder formalism has
been introduced firstly by Kadanoff and Ceva \cite{ceva} in order to discuss
the existence of a generalized statistics. Posteriorly this was extended to
the continuum quantum field theory \cite{swieca}. This procedure has been
applied to some models in (2+1) dimensions by using a new interpretation of
the operators that generate the statistics. Now, over the plane ($%
x^{1},\,x^{2}$), the Maxwell theory has a nontrivial value for the
topological charge associated with the identically conserved current $J^{\mu
}\,=\,\epsilon ^{\mu \nu \rho }\partial _{\nu }A_{\rho }$. The magnetic flux
content correspondent to $J^{\mu }$ is described by a non-local operator
(vortex operator) $\mu (x)$ defined on a certain curve $C$. The correlation
function $<\mu (1)\mu (2)>$ of the disorder operator is given as Euclidean
functional integrals. In the same way, we can define the charge bearing
operator $\sigma (x)$ as being a dual version of $\mu $.

In order to give a better understanding of the role of order-disorder
correlation functions, we will take as example the case of the
Maxwell-Chern-Simons theory, since its photon propagator in the coordinate
space will be useful in the follow.

The order correlation function for the MCS theory is defined in terms of the
following Euclidean functional integral 
\begin{equation}
<\sigma \left( x\right) \,\sigma *(y)>\,=Z^{-1}\,\int DA_{\mu }\,\exp
\left\{ -\int d^{3}z\,\left[ \frac{1}{2}A_{\mu }\left( P^{\mu \nu }+\theta
\,C^{\mu \nu }+G^{\mu \nu }\right) A_{\nu }+\,C_{\mu }\,A^{\mu }\right]
\right\}   \label{eq14}
\end{equation}

\noindent where $P^{\mu \nu }\equiv \,-\,\Box \,\delta ^{\mu \nu
}\,+\,\partial ^{\mu }\partial ^{\nu }$, $C^{\mu \nu }\equiv \,-i\,\epsilon
^{\mu \alpha \nu }\partial _{\alpha }$ and $G^{\mu \nu }$ is the usual gauge
fixing term. Here we adopt an external field $C_{\mu }$.

Integrating over $A_{\mu }$ we readily obtain 
\begin{equation}
<\sigma \left( x\right) \,\sigma *(y)>\,=\,\exp \left\{ \frac{1}{2}\int
d^{3}z\,d^{3}z\,^{\prime }\,C_{\mu }\left( z\right) \left[ P^{\mu \nu
}+\theta \,C^{\mu \nu }+G^{\mu \nu }\right] ^{-1}C_{\nu }\left( z\,^{\prime
}\right) \right\}   \label{eq15}
\end{equation}

\noindent with $\left[ P^{\mu \nu }+\theta \,C^{\mu \nu }+G^{\mu \nu
}\right] ^{-1}=\,<A_{\mu }(x)\,A_{\nu }(y)>_{MCS}$ being the Euclidean
propagator of the $A_{\mu }$ field in MCS theory. Its explicit expression in
the coordinate space is given by 
\begin{equation}
<A_{\mu }(x)\,A_{\nu }(y)>_{MCS}=\,\left[ P^{\mu \nu }+i\,\theta \,\epsilon
^{\mu \alpha \nu }\partial _{\alpha }\right] \left[ \frac{1-e^{-\theta \,R}}{%
4\pi \theta ^{2}R}\right] \,-\,\lim_{m\rightarrow 0}\,\xi \,\,\partial ^{\mu
}\,\partial ^{\nu }\left[ \frac{1}{m}-\,\frac{R}{8\pi }\right]   \label{eq16}
\end{equation}

\noindent Before going on, we should remark that $<\sigma \left( x\right)
\,\sigma *(y)>\,$is not a gauge invariant quantity. The reason is that under
a formal gauge transformation $A_{\mu }\,\rightarrow \,A_{\mu }\,+\,\Lambda $%
, the charge operator changes to $\sigma ^{\prime }\,=\,\exp \left( 2\pi
\,i\,\Lambda \left( x\right) \right) \,\sigma $. In this way, going back to
equation (\ref{eq15}), we must extract the gauge independent part of $%
<\sigma \left( x\right) \,\sigma *(y)>$. This will be achieved by inserting
the gauge independent part of $<A_{\mu }(x)\,A_{\nu }^{*}(y)>_{MCS}$, namely 
$\delta ^{\mu \nu }$ and $\epsilon ^{\mu \alpha \nu }$ proportional terms.
At the end of the calculations, it can be shown that only the diagonal part
of $<A_{\mu }(x)\,A_{\nu }^{*}(y)>_{MCS}$ proportional to $\delta ^{\mu \nu
}\Box $, contribute to the order correlation function. Therefore we obtain
the following expression 
\begin{equation}
<\sigma \left( x\right) \,\sigma *(y)>_{MCS}=\,\exp \left\{ <A_{\mu
}(x)\,A_{\nu }^{*}(y)>_{\xi =0}^{diag}\right\} \,\exp \left( \frac{%
e^{-\theta \,R}}{4\pi R}\right) =  \label{eq17}
\end{equation}

\begin{equation}
=\,\exp \left\{ \frac{\pi \,a^{2}}{\bar{\theta}}\left[ e^{-\,\bar{\theta}%
\,R}\,-\,1\right] \right\} \Longrightarrow <\sigma _{R}\left( x\right)
\,\sigma _{R}*(y)>\,=\,\exp \left\{ \frac{\pi \,a^{2}}{\bar{\theta}}\,e^{-\,%
\bar{\theta}\,R}\right\}   \label{eq18}
\end{equation}

\noindent where it was adopted the renormalization $\sigma _{R}\,\equiv
\,\sigma \,e^{\frac{\pi \,a^{2}}{2\,\bar{\theta}}}$ and $a$ is a charge
parameter. As a consequence, $\lim_{R\rightarrow \infty }<\sigma _{R}\left(
x\right) \,\sigma _{R}*(y)>=1$, which reflects the screening of the charge
associated with the mass generation for the gauge field induced by the CS
term.

Now, going back to our model, we begin considering the limit which exclude
the Podolsky term, $\bar{\theta}^{2}\gg \alpha ^{2}$, in equation (\ref{eq13}%
) 
\begin{equation}
<A_{\mu }\left( x\right) \,A_{\nu }(y)>_{\xi =0}^{dual}\,\cong \,\left[
P_{\mu \nu }\,+\,2\,i\,\bar{\theta}\,\epsilon _{\mu \nu \alpha }\,\partial
^{\alpha }\right] \left( \frac{1}{\bar{\theta}^{2}}\,-\,1\right) \left( 
\frac{4}{\bar{\theta}^{2}}\right) \frac{e^{-\,\bar{\theta}\,R}}{4\,\pi \,R},
\label{eq19}
\end{equation}

\noindent in order to compare it with results of the MCS case. By examining
the diagonal part of the above propagator we have 
\[
<A_{\mu }\left( x\right) \,A_{\nu }(y)>^{dual}\,\,\cong \,\Box \,\delta
_{\mu \nu }\left( \frac{1}{\bar{\theta}^{2}}\,-\,1\right) \left( \frac{4}{%
\bar{\theta}^{2}}\right) \frac{e^{-\,\bar{\theta}\,R}}{4\,\pi \,R}\,\cong
\,\left( \frac{1}{\bar{\theta}^{2}}\,-\,1\right) \frac{e^{-\,\bar{\theta}\,R}%
}{4\,\pi \,R}\,+
\]
\begin{equation}
+\,extra\,\,terms\,=\,-\,\left( \frac{1}{\bar{\theta}^{2}}\,-\,1\right)
<A_{\mu }\left( x\right) \,A_{\nu }(y)>_{MCS}  \label{eq20}
\end{equation}

\noindent where ``extra terms'' are proportional to the $\delta $-functions.
From equations (\ref{eq16}) and (\ref{eq18}) we expect that 
\begin{equation}
<\sigma _{R}\left( x\right) \,\sigma _{R}*(y)>^{dual}\,\cong \,\exp \left[
\left( \frac{1}{\bar{\theta}^{2}}\,-\,1\right) ^{-1}\right] <\sigma
_{R}\left( x\right) \,\sigma _{R}*(y)>_{MCS},  \label{eq21}
\end{equation}

\noindent for $R\longrightarrow \infty $%
\begin{equation}
<\sigma _{R}\left( x\right) \,\sigma _{R}*(y)>^{dual}\,\longrightarrow
\,const.  \label{eq22}
\end{equation}

\noindent Therefore the order correlation function, which is associated with
charge screening, in our model has a similar behavior to that of the MCS
theory. The result given by equation (\ref{eq19}) express the charge
screening, which in this case is 
\begin{equation}
Q^{dual}\,=\,\int d^{2}z\,J^{0}\,=\,\theta \,\int d^{2}z\,\nabla _{\xi
}^{2}\,\epsilon _{ij}\,\partial _{z}^{i}\,A^{i}(z,\xi ).  \label{eq23}
\end{equation}

\noindent which differs from the usual MCS charge by a second order
derivative operator. Here $J_{\mu }\,\equiv \,\partial ^{\nu }J_{\mu \nu }$
defined in equation (\ref{eq24}) below. Note that the presence of the
differential operators in $Q^{dual}$ do not alter the long range distance
behavior of the order correlation function when compared with MCS theory.

Now, in order to build the disorder correlation function $<\mu \left(
1\right) \mu \left( 2\right) >$ in our model, we begin defining the vortex
operator which is associated to the magnetic flux on the plane $\left(
x^{1},\,x^{2}\right) $. This is obtained by coupling a certain external
field $W_{\mu }$ to the dual current through

\begin{equation}
J_{\theta }^{\mu \nu }\,\equiv \,F^{\mu \nu }\,-\,\frac{\theta \,\Box
\,\epsilon ^{\mu \nu \alpha }}{\left( 1\,-\,4\,\alpha ^{2}\,\Box \right) }%
\,A_{\alpha },  \label{eq24}
\end{equation}

\noindent which comes from the equation of motion. The generalized disorder
operator can be written as 
\begin{equation}
\mu _{\theta }^{dual}(x)\,=\,\exp \left\{ -i\,b\,\int d^{3}z\,J_{\theta
}^{\mu \nu }\,W_{\mu \nu }\right\} ,  \label{eq25}
\end{equation}

\noindent where $W^{\mu \nu }$ is an external tensor field, $W_{\mu \nu
}\,\equiv \,\partial _{\mu }\,W_{\nu }\,-\,\partial _{\nu }\,W_{\mu }$,
which would be coupled to the conserved current $J_{\theta }^{\mu \nu }$ in
order to obtain the correct correlation function 
\[
<\mu _{\theta }\left( x\right) \,\mu *_{\theta }(y)>^{dual}\,=\,\int DA_{\mu
}\exp \left\{ -\,\int d^{3}z\,\left[ \frac{1}{2}A^{\mu }D_{\mu \nu
}^{dual}\,A^{\nu }\,+\right. \right. ,
\]
\begin{equation}
\left. \,\left. +\,\,A^{\mu }D_{\mu \nu (GCS)}\,W^{\nu }\,+\,\frac{1}{4}%
\left( W_{\mu \nu }\right) ^{2}\right] \right\}   \label{eq26}
\end{equation}

\noindent with $GCS$ standing for Generalized Chern-Simons, and $D_{\mu \nu
}^{dual}$ is given by 
\[
D_{\mu \nu }^{dual}\,\equiv \,\left( 1\,-\,4\,\alpha ^{2}\,\Box \right)
P_{\mu \nu }\,-\,\xi \,\partial _{\mu }\,\partial _{\nu }\,-\,\theta
\,\epsilon _{\mu \alpha \nu }\,\partial ^{\alpha }\Box 
\]
\begin{equation}
D_{\mu \nu }^{GCS}\,\equiv \,P_{\mu \nu }\,-\,\theta \,\epsilon _{\alpha \nu
}\,\partial ^{\alpha }\,\Box \,\left( 1\,-\,4\,\alpha ^{2}\,\Box \right)
^{-1}  \label{eq27}
\end{equation}

\noindent where $P_{\mu \nu }\,\equiv \,-\,\Box \,\delta _{\mu \nu
}\,+\,\partial _{\mu }\,\partial _{\nu }$. Now, if we consider the action of
the operators $P_{\mu \nu }$ and $\theta \epsilon _{\mu \alpha \nu
}\,\partial ^{\alpha }\Box $ over $W_{\mu \nu }$, gives rise 
\[
P_{\mu \nu }\,W^{\mu }\,\longrightarrow \,Z_{\nu }\,=\,\int \,d\lambda _{\nu
}\,\delta ^{3}\left( z-\lambda \right) 
\]
\begin{equation}
\frac{\theta \,\Box \,\epsilon _{\mu \alpha \nu }\,\partial ^{\alpha
}\,W^{\mu }}{\left( 1\,-\,4\,\alpha ^{2}\,\Box \right) }\,\longrightarrow
\,U_{\nu }\,=\,\theta \,\int \,d\lambda ^{\mu }\,\epsilon _{\mu \alpha \nu
}\,\partial ^{\alpha }\,\Box \,(1\,-\,4\,\alpha ^{2}\,\Box )^{-1}\delta
^{3}\left( z-\lambda \right) .  \label{eq28}
\end{equation}

\noindent Therefore, after integration over the field $A_{\mu }$ in equation
(\ref{eq23}) we get 
\[
<\mu _{\theta }\left( x\right) \,\mu *_{\theta }(y)>_{N=1}^{dual}\,=\,\exp
\left\{ \frac{1}{2}\int d^{3}z\,d^{3}\acute{z}\,\left( Z_{\mu
}(z,x,y)\,+\,U_{\mu }(z,x,y)\right) \times \right. 
\]
\begin{equation}
\left. \times \,<A_{\mu }(x)\,A*_{\nu }(y)>^{dual}\,\left( Z_{\nu }(\acute{z}%
,x,y)\,+\,U_{\nu }(\acute{z},x,y)\right) \,-\,\frac{1}{4}\left( W_{\mu \nu
}\right) ^{2}\right\} ,  \label{eq29}
\end{equation}

\noindent where $<A_{\mu }(x)\,A*_{\nu }(y)>^{dual}$ is given by equation(%
\ref{eq13}). Now, if we now turn our attention to the fact that in the limit 
$\theta \gg 4\,\alpha ^{2}$, the photon correlation function is given by
equation (\ref{eq19}) and the field $U_{\nu }$ will not depend on the factor 
$4\,\alpha ^{2}$, we will have 
\[
<\mu _{\theta }\left( x\right) \,\mu *_{\theta }(y)>_{N=1}^{dual}\,=\,\exp
\left\{ \frac{1}{2}\left( 1-\frac{1}{\bar{\theta}^{2}}\right) \int
d^{3}z\,d^{3}\acute{z}\left( Z_{\mu }\,+\,\tilde{U}_{\mu }\right) \right. 
\]
\begin{equation}
\left. \,<A_{\mu }\,A*_{\nu }>^{MCS}\,\left( Z_{\nu }\,+\,\tilde{U}_{\nu
}\right) \,-\,\frac{1}{4}\left( W_{\mu \nu }\right) ^{2}\right\} ,
\label{eq30}
\end{equation}

\noindent with $\tilde U_\nu \,=\,\theta \,\int_{x,L}d\lambda ^\mu
\,\epsilon _{\mu \alpha \nu }\,\partial ^\alpha \,\Box \,\delta ^3(z-\lambda
)$.

Now, we note that up to $\Box $ term into $\tilde{U}_{\nu }$ field, the
integrand of the above equation corresponds to the correlation function of
MCS theory. However, since $<A_{\mu }(x)\,A*_{\nu }(y)>_{MCS}$ depends on $%
1/\left| z-\acute{z}\right| $ the contractions which involve the field $%
\tilde{U}_{\nu }$ give rise to delta functions $\delta ^{3}(z-\lambda )$ and 
$\delta ^{3}(z-\eta )$ such that the line integral over $d\lambda _{\mu }$
and $d\eta _{\mu }$ vanishes. This means that the integrand of the equation
(30) corresponds to that of the MCS theory or, 
\begin{equation}
<\mu _{\theta }\left( x\right) \,\mu *_{\theta }(y)>^{dual}\,=\,\exp
{}^{\left( 1-\frac{1}{\theta ^{2}}\right) }<\mu _{\theta }\left( x\right)
\,\mu *_{\theta }(y)>^{MCS}.  \label{eq31}
\end{equation}

\noindent Therefore, since the behavior of the vortex correlation function
operator in the MCS theory for very large distances $\left| x-y\right|
\longrightarrow \infty $ is a constant, indicating that $\mu _{\theta }$
does not create genuine vortex excitations, we expect the same behavior for
the dual theory.

For a future program, we intend to investigate the possible connection with
the interesting formalism developed by Barci et al, where it was made a
mapping among some models in three dimensions \cite{uerj1}. This was done by
using a nonlinear redefinition of the gauge field, in contrast to the linear
self-dual transformation used in this work.

{\bf Acknowledgments: }The authors are grateful to CNPq and FAPESP for
partial financial support, and to D. Dalmazi for discussions.

\end{document}